\documentclass{article}\usepackage{spconf}\usepackage[stretch=50,shrink=50]{microtype}

\usepackage{epsfig,amsmath,amssymb,epsf,amsthm,scalefnt,multirow,bm,mathtools,stfloats,stmaryrd}
\usepackage{xcolor}
\usepackage{float}
\usepackage{cite}
\usepackage{psfrag}
\usepackage{algorithm}
\usepackage{algpseudocode}

\newtheorem*{lemma*}{Lemma}

\DeclareMathOperator*{\argmax}{argmax}

\algrenewcommand\algorithmicrequire{\textbf{Input:}}
\algrenewcommand\algorithmicensure{\textbf{Output:}}
\algdef{SE}[DOWHILE]{Do}{doWhile}{\algorithmicdo}[1]{\algorithmicwhile\ #1}

\begin{document}

\title{Gridless Channel Estimation for MmWave Hybrid Massive MIMO Systems with Low-Resolution ADCs}

\name{In-soo Kim and Junil Choi
\thanks{This work was supported in part by the National Research Foundation (NRF) grant funded by the MSIT of the Korean Government under Grant 2019R1C1C1003638, in part by the Ministry of Science and ICT (MSIT), South Korea, under the Information Technology Research Center (ITRC) support program supervised by the Institute of Information and Communications Technology Planning and Evaluation (IITP) under Grant IITP-2020-0-01787, and in part by the Institute of Information and Communications Technology Planning and Evaluation (IITP) grant funded by the Korean Government (MSIT) (Development of Integer-Forcing MIMO Transceivers for 5G and Beyond Mobile Communication Systems) under Grant 2016-0-00123.}}
\address{School of Electical Engineering\\
KAIST\\
Daejeon, Korea\\
E-mail: \{insookim, junil\}@kaist.ac.kr}

\maketitle

\begin{abstract}
This paper proposes the Newtonized fully corrective forward greedy selection-cross validation-based (NFCFGS-CV-based) channel estimator for millimeter (mmWave) hybrid massive multiple-input multiple-output (MIMO) systems with low-resolution analog-to-digital converters (ADCs). The proposed NFCFGS algorithm is a gridless compressed sensing (CS) technique that combines the FCFGS and Newtonized orthogonal matching pursuit (NOMP) algorithms. In particular, NFCFGS performs single path estimation over the continuum at each iteration based on the previously estimated paths. The CV technique is adopted as an indicator of termination in the absence of the prior knowledge on the number of paths, which is a model validation technique that prevents overfitting.
\end{abstract}


\section{Introduction}
The millimeter (mmWave) band in the range of $30$-$300$ GHz and massive multiple-input multiple-output (MIMO) are expected to significantly enhance the performance of future wireless communications systems \cite{6894453}. The high sampling frequency and large number of radio frequency (RF) chains, however, result in a prohibitive amount of power consumption. Among various solutions, hybrid architectures and low-resolution analog-to-digital converters (ADCs) are considered as practical candidates that achieve high energy efficiency \cite{7876856}.

To perform sparse recovery in massive MIMO systems with low-resolution ADCs, various compressed sensing (CS) algorithms were proposed \cite{8171203, 8320852, 8310593, 8515242}. The generalized approximate message passing (GAMP) \cite{8171203}, vector AMP (VAMP) \cite{8171203}, generalized expectation consistent-signal recovery (GEC-SR) \cite{8320852}, and variational Bayesian-sparse Bayesian learning (VB-SBL) \cite{8310593} algorithms are CS techniques that estimate the parameters of interest on the grid, so the off-grid error is inevitable. The atomic norm minimization (ANM) algorithm for mixed one-bit ADCs \cite{8515242}, on the other hand, estimates the channel off the grid. ANM, however, requires high-resolution ADCs to determine the regularization parameter that is essential for ANM to run.

In this paper, the Newtonized fully corrective forward greedy selection-cross validation-based (NFCFGS-CV-based) channel estimator is proposed for mmWave hybrid massive MIMO systems with low-resolution ADCs. The proposed NFCFGS algorithm is a gridless CS technique that is inspired by the FCFGS \cite{doi:10.1137/090759574} and Newtonized orthogonal matching pursuit (NOMP) \cite{7491265} algorithms. NFCFGS estimates single path off the grid at each iteration based on the previously estimated paths. To determine when to terminate without any prior knowledge on the number of paths, the CV technique is adopted, which is a model validation technique for overfitting prevention.

\textbf{Notation:} $a$, $\mathbf{a}$, and $\mathbf{A}$ denote a scalar, vector, and matrix. The element restriction of $\mathbf{a}$ to the index set $\mathcal{I}$ is $\mathbf{a}_{\mathcal{I}}$. $\mathbf{a}\succ\mathbf{b}$ represents the element-wise inequality between $\mathbf{a}$ and $\mathbf{b}$. The column-wise vectorization of $\mathbf{A}$ is $\mathrm{vec}(\mathbf{A})$. The Kronecker product of $\mathbf{A}$ and $\mathbf{B}$ is $\mathbf{A}\otimes\mathbf{B}$. $\llbracket n\rrbracket$ denotes $\llbracket n\rrbracket=\{1, \dots, n\}$.

\section{System Model}
Consider an uplink massive MIMO system with a base station and $K$ single-antenna users. The base station has $M$ antennas and $R$ RF chains where $R\leq M$. As a means to reduce the power consumption, each RF chain is equipped with a pair of $B$-bit ADCs for the real and imaginary parts. The system operates in the mmWave wideband with $D$ delay taps and sampling period $T_{s}$. In this section, the system model of the channel estimation phase is described.

The received signal $\mathbf{y}[n]\in\mathbb{C}^{R}$ at time $n$ is
\begin{align}\label{received_signal_at_n}
\mathbf{y}[n]=&\mathbf{W}[n]^{\mathrm{H}}\times\notag\\
              &\left(\underbrace{\begin{bmatrix}\mathbf{H}[0]&\cdots&\mathbf{H}[D-1]\end{bmatrix}}_{=\mathbf{H}}\underbrace{\begin{bmatrix}\mathbf{s}[n]\\\vdots\\\mathbf{s}[n-D+1]\end{bmatrix}}_{=\mathbf{s}_{n}}+\bar{\mathbf{v}}[n]\right)\notag\\
             =&\mathbf{W}[n]^{\mathrm{H}}\mathbf{H}\mathbf{s}_{n}+\mathbf{W}[n]^{\mathrm{H}}\bar{\mathbf{v}}[n]\notag\\
             =&(\underbrace{\mathbf{s}_{n}^{\mathrm{T}}\otimes\mathbf{W}[n]^{\mathrm{H}}}_{=\bar{\mathbf{A}}[n]})\underbrace{\mathrm{vec}(\mathbf{H})}_{=\mathbf{h}}+\underbrace{\mathbf{W}[n]^{\mathrm{H}}\bar{\mathbf{v}}[n]}_{=\mathbf{v}[n]}
\end{align}
where the last line is due to $\mathrm{vec}(\mathbf{ABC})=(\mathbf{C}^{\mathrm{T}}\otimes\mathbf{A})\mathrm{vec}(\mathbf{B})$. Here, $\mathbf{H}[d]\in\mathbb{C}^{M\times K}$ is the $d$-th delay tap, $\mathbf{s}[n]\in\mathbb{C}^{K}$ is the training signal at time $n$ under the transmit power constraint $\mathbb{E}\{\mathbf{s}[n]\mathbf{s}[n]^{\mathrm{H}}\}=\rho\mathbf{I}_{K}$, $\bar{\mathbf{v}}[n]\sim\mathcal{CN}(\mathbf{0}_{M}, \mathbf{I}_{M})$ is the additive white Gaussian noise (AWGN) at time $n$, and $\mathbf{W}[n]\in\mathbb{C}^{M\times R}$ is the RF combiner at time $n$ with a network of phase shifters that satisfies $\mathbf{W}[n]^{\mathrm{H}}\mathbf{W}[n]=\mathbf{I}_{R}$. Therefore, $\mathbf{v}[n]\sim\mathcal{CN}(\mathbf{0}_{R}, \mathbf{I}_{R})$, and the signal-to-noise ratio (SNR) is $\rho$.

The channels in the mmWave band consist of a small number of paths due to the severe path loss. Therefore, the channel of the $k$-th user that corresponds to the $k$-th column of $\mathbf{H}[d]$ is \cite{6834753}
\begin{align}\label{channel}
\mathbf{h}_{k}[d]&=\sum_{\ell=1}^{L_{k}}\alpha_{k, \ell}\underbrace{p(dT_{s}-\tau_{k, \ell})\mathbf{a}(\theta_{k, \ell})}_{=\mathbf{a}_{d}(\theta_{k, \ell}, \tau_{k, \ell})}\notag\\
                 &=\underbrace{\begin{bmatrix}\mathbf{a}_{d}(\theta_{k, 1}, \tau_{k, 1})&\cdots&\mathbf{a}_{d}(\theta_{k, L_{k}}, \tau_{k, L_{k}})\end{bmatrix}}_{=\mathbf{F}_{k}[d]}\underbrace{\begin{bmatrix}\alpha_{k, 1}\\\vdots\\\alpha_{k, L_{k}}\end{bmatrix}}_{=\bm{\alpha}_{k}}
\end{align}
where $L_{k}$ is the number of paths, $\alpha_{k, \ell}\sim\mathcal{CN}(0, 1)$ is the $\ell$-th path gain, $\theta_{k, \ell}\sim\mathrm{Uniform}([-\pi/2, \pi/2])$ is the $\ell$-th angle-of-arrival (AoA), $\tau_{k, \ell}\sim\mathrm{Uniform}([0, (D-1)T_{s}])$ is the $\ell$-th delay, $\mathbf{a}(\theta)\in\mathbb{C}^{M}$ is the array response vector, and $p(t)$ is the pulse shaping filter.

The received signal over the channel estimation phase of length $N$ is
\begin{equation}\label{received_signal_over_N}
\mathbf{y}=\begin{bmatrix}\mathbf{y}[1]\\\vdots\\\mathbf{y}[N]\end{bmatrix}=\underbrace{\begin{bmatrix}\bar{\mathbf{A}}[1]\\\vdots\\\bar{\mathbf{A}}[N]\end{bmatrix}}_{=\bar{\mathbf{A}}}\mathbf{h}+\underbrace{\begin{bmatrix}\mathbf{v}[1]\\\vdots\\\mathbf{v}[N]\end{bmatrix}}_{=\mathbf{v}}.
\end{equation}
Then, $\mathbf{y}$ is quantized by $B$-bit ADCs as $\hat{\mathbf{y}}=\mathrm{Q}(\mathbf{y})$ where $\mathrm{Q}(\cdot)$ is the $B$-bit quantization function that operates element-wise as
\begin{equation}\label{quantization_function}
\hat{y}=\mathrm{Q}(y)\iff\begin{cases}\mathrm{Re}(\hat{y}^{\mathrm{lo}})\leq\mathrm{Re}(y)\leq\mathrm{Re}(\hat{y}^{\mathrm{up}})\\\mathrm{Im}(\hat{y}^{\mathrm{lo}})\leq\mathrm{Im}(y)\leq\mathrm{Im}(\hat{y}^{\mathrm{up}})\end{cases}.
\end{equation}
Here, $\hat{y}^{\mathrm{lo}}\in\mathbb{C}$ and $\hat{y}^{\mathrm{up}}\in\mathbb{C}$ are the lower and upper thresholds that map to $\hat{y}\in\mathbb{C}$, so the real and imaginary parts of $\hat{y}^{\mathrm{lo}}$, $\hat{y}^{\mathrm{up}}$, and $\hat{y}$ correspond to one of the $2^{B}$ quantization intervals. For notational convenience, the collection of the lower and upper thresholds that map to $\hat{\mathbf{y}}$ are denoted as $\hat{\mathbf{y}}^{\mathrm{lo}}\in\mathbb{C}^{RN}$ and $\hat{\mathbf{y}}^{\mathrm{up}}\in\mathbb{C}^{RN}$.

Now, let us rewrite $\mathbf{h}$ as
\begin{equation}
\mathbf{h}=\underbrace{\begin{bmatrix}\mathbf{F}_{1}[0]&&\\&\ddots&\\&&\mathbf{F}_{K}[0]\\&\vdots&\\\mathbf{F}_{1}[D-1]&&\\&\ddots&\\&&\mathbf{F}_{K}[D-1]\end{bmatrix}}_{=\mathbf{F}(\mathcal{P})}\underbrace{\begin{bmatrix}\bm{\alpha}_{1}\\\vdots\\\bm{\alpha}_{K}\end{bmatrix}}_{=\bm{\alpha}}
\end{equation}
using \eqref{received_signal_at_n} and \eqref{channel}. Here, $\mathcal{P}$ is the collection of all $(\theta_{k, \ell}, \tau_{k, \ell})$. Then, \eqref{received_signal_over_N} becomes
\begin{equation}\label{modified_received_signal_over_N}
\mathbf{y}=\bar{\mathbf{A}}\mathbf{F}(\mathcal{P})\bm{\alpha}+\mathbf{v}
\end{equation}
where $\mathbf{A}(\mathcal{P})=\bar{\mathbf{A}}\mathbf{F}(\mathcal{P})\in\mathbb{C}^{RN\times(L_{1}+\cdots+L_{K})}$. The goal is to estimate $(\bm{\alpha}, \mathcal{P})$ from $\hat{\mathbf{y}}$. The problem, however, is that $\mathbf{F}(\mathcal{P})$ is nonlinear with respect to $\mathcal{P}$. Furthermore, there is no prior knowledge on the number of paths, which further complicates the situation.

\section{Proposed NFCFGS-CV Algorithm}
In this section, the NFCFGS-CV-based channel estimator is proposed. The NFCFGS algorithm is a gridless CS technique that performs single path estimation at each iteration. The termination condition of NFCFGS is determined by the CV technique, which is a model validation technique that prevents overfitting.

\subsection{Proposed NFCFGS Algorithm}
The proposed NFCFGS algorithm is a gridless CS technique that combines FCFGS \cite{doi:10.1137/090759574} and the NOMP algorithm \cite{7491265}. In particular, one iteration of NFCFGS consists of single path estimation that is performed based on the previously estimated paths. To develop NFCFGS, define $\hat{\mathcal{P}}$ as the collection of all the previously estimated $(\theta_{k, \ell}, \tau_{k, \ell})$, and $\hat{\bm{\alpha}}\in\mathbb{C}^{|\hat{\mathcal{P}}|}$ as the collection of all the previously estimated $\alpha_{k, \ell}$. The goal is to estimate the path that is parametrized by $(\alpha, \theta, \tau, k)$ using $(\hat{\bm{\alpha}}, \hat{\mathcal{P}})$, while $k\in\llbracket K\rrbracket$ is unknown a priori.

To proceed, define $\mathbf{a}(\theta, \tau, k)\in\mathbb{C}^{RN}$ using $(\theta, \tau, k)$ by the same logic as $\mathbf{A}(\mathcal{P})$ is defined in \eqref{modified_received_signal_over_N} using $\mathcal{P}$. Then, the unquantized received signal becomes
\begin{equation}
\mathbf{y}=\alpha\mathbf{a}(\theta, \tau, k)+\mathbf{A}(\hat{\mathcal{P}})\hat{\bm{\alpha}}+\mathbf{v}
\end{equation}
where $\mathbf{z}=\mathbf{A}(\hat{\mathcal{P}})\hat{\bm{\alpha}}\in\mathbb{C}^{RN}$, and the likelihood function is
\begin{align}\label{likelihood_function}
&\ell_{\mathbf{z}}(\alpha, \theta, \tau, k)=\Pr\left[\hat{\mathbf{y}}\middle|\alpha, \theta, \tau, k, \hat{\bm{\alpha}}, \hat{\mathcal{P}}\right]=\notag\\
&\Pr\left[\mathrm{Re}(\hat{\mathbf{y}}^{\mathrm{lo}})\preceq\mathrm{Re}(\mathbf{y})\preceq\mathrm{Re}(\hat{\mathbf{y}}^{\mathrm{up}})\middle|\alpha, \theta, \tau, k, \hat{\bm{\alpha}}, \hat{\mathcal{P}}\right]\times\notag\\
&\Pr\left[\mathrm{Im}(\hat{\mathbf{y}}^{\mathrm{lo}})\preceq\mathrm{Im}(\mathbf{y})\preceq\mathrm{Im}(\hat{\mathbf{y}}^{\mathrm{up}})\middle|\alpha, \theta, \tau, k, \hat{\bm{\alpha}}, \hat{\mathcal{P}}\right]
\end{align}
where the last equality is due to $\eqref{quantization_function}$ and the fact that the real and imaginary parts of $\mathbf{v}\sim\mathcal{CN}(\mathbf{0}_{RN}, \mathbf{I}_{RN})$ are independent. Since
\begin{gather}
\mathrm{Re}(\hat{\mathbf{y}}^{\mathrm{lo}})\preceq\mathrm{Re}(\mathbf{y})\preceq\mathrm{Re}(\hat{\mathbf{y}}^{\mathrm{up}})\notag\\
\Updownarrow\notag\\
\mathrm{Re}(\hat{\mathbf{y}}^{\mathrm{lo}}-\mathbf{z})\preceq\mathrm{Re}(\alpha\mathbf{a}(\theta, \tau, k)+\mathbf{v})\preceq\mathrm{Re}(\hat{\mathbf{y}}^{\mathrm{up}}-\mathbf{z})\notag,
\end{gather}
the first term in the last equality of \eqref{likelihood_function} corresponds to the probability that $\mathcal{N}(\mathrm{Re}(\alpha\mathbf{a}(\theta, \tau, k)), 1/2\cdot\mathbf{I}_{RN})$ be in the box that has $\mathrm{Re}(\hat{\mathbf{y}}^{\mathrm{lo}}-\mathbf{z})$ and $\mathrm{Re}(\hat{\mathbf{y}}^{\mathrm{up}}-\mathbf{z})$ as the lower and upper boundaries. The same argument holds for the imaginary part. The two terms of $\ell_{\mathbf{z}}(\alpha, \theta, \tau, k)$ have well-known closed-form expressions that are log-concave with respect to $\alpha$ \cite{9351751, 7439790, boyd2004convex}. Therefore, $\ell_{\mathbf{z}}(\alpha, \theta, \tau, k)$ is log-concave with respect to $\alpha$. The problem, however, is that $\ell_{\mathbf{z}}(\alpha, \theta, \tau, k)$ is nonconvex with respect to $(\theta, \tau, k)$.

In analogy to FCFGS that performs support recovery by identifying the support that maximizes the derivative of the log-likelihood function with respect to the support element at $0$, NFCFGS estimates $(\theta, \tau, k)$ as follows. First, NFCFGS interprets $(\theta, \tau, k)$ and $\alpha$ as the support and support element. Then, $(\theta, \tau, k)$ is estimated by solving
\begin{alignat}{2}\label{aoa_delay_estimation}
&\underset{\theta, \tau, k}{\text{maximize}}\ &&\underbrace{|\nabla_{\alpha}\log \ell_{\mathbf{z}}(0, \theta, \tau, k)|^{2}}_{=f_{\mathbf{z}}(\theta, \tau, k)}\notag\\
&\text{subject}\ \text{to}\                   &&(\theta, \tau, k)\in[-\pi/2, \pi/2]\times[0, (D-1)T_{s}]\times\llbracket K\rrbracket.
\end{alignat}

To solve \eqref{aoa_delay_estimation}, NFCFGS first maximizes $f_{\mathbf{z}}(\theta, \tau, k)$ on the grid
\begin{equation}
\Omega\subseteq[-\pi/2, \pi/2]\times[0, (D-1)T_{s}]\times\llbracket K\rrbracket
\end{equation}
that discretizes the constraint in \eqref{aoa_delay_estimation}. In practice, $[-\pi/2, \pi/2]$ and $[0, (D-1)T_{s}]$ are discretized by $\geq 2M$ and $\geq 2D$ points \cite{7491265, 7961152, 7458188}. Then, the coarse estimate $(\hat{\theta}, \hat{\tau})$ is refined off the grid using Newton's method for nonconvex optimization as \cite{murphy2012machine}
\begin{align}
(\hat{\theta}, \hat{\tau})+\begin{cases}\eta\mathbf{n}_{\mathbf{z}}(\hat{\theta}, \hat{\tau}, \hat{k})&\text{if}\ \nabla_{(\theta, \tau)}^{2}f_{\mathbf{z}}(\hat{\theta}, \hat{\tau}, \hat{k})\prec\mathbf{0}_{2\times 2}\\
                                        \eta\mathbf{g}_{\mathbf{z}}(\hat{\theta}, \hat{\tau}, \hat{k})&\text{if}\ \nabla_{(\theta, \tau)}^{2}f_{\mathbf{z}}(\hat{\theta}, \hat{\tau}, \hat{k})\nprec\mathbf{0}_{2\times 2}\end{cases}
\end{align}
at each iteration where
\begin{align}
&\mathbf{n}_{\mathbf{z}}(\theta, \tau, k)=-\nabla_{(\theta, \tau)}^{2}f_{\mathbf{z}}(\theta, \tau, k)^{-1}\nabla_{(\theta, \tau)}f_{\mathbf{z}}(\theta, \tau, k),\\
&\mathbf{g}_{\mathbf{z}}(\theta, \tau, k)=\nabla_{(\theta, \tau)}f_{\mathbf{z}}(\theta, \tau, k)
\end{align}
are the Newton and gradient steps, and $\eta$ is the step size. The objective of the Newton refinement is to reduce the off-grid error, which is a NOMP-inspired technique. Finally, NFCFGS updates $\hat{\mathcal{P}}$ as $\hat{\mathcal{P}}\cup\{(\hat{\theta}, \hat{\tau}, \hat{k})\}$.

To update the path gains, denote the collected path gains of $\hat{\mathcal{P}}$ that are to be estimated as $\mathbf{x}\in\mathbb{C}^{|\hat{\mathcal{P}}|}$. Then, the unquantized received signal is $\mathbf{y}=\mathbf{A}(\hat{\mathcal{P}})\mathbf{x}+\mathbf{v}$, and the likelihood function is given as
\begin{align}
&\ell(\mathbf{x}, \hat{\mathcal{P}})=\Pr\left[\hat{\mathbf{y}}\middle|\mathbf{x}, \hat{\mathcal{P}}\right]=\notag\\
&\Pr\left[\mathrm{Re}(\hat{\mathbf{y}}^{\mathrm{lo}})\preceq\mathrm{Re}(\mathbf{A}(\hat{\mathcal{P}})\mathbf{x}+\mathbf{v})\preceq\mathrm{Re}(\hat{\mathbf{y}}^{\mathrm{up}})\middle|\mathbf{x}, \hat{\mathcal{P}}\right]\times\notag\\
&\Pr\left[\mathrm{Im}(\hat{\mathbf{y}}^{\mathrm{lo}})\preceq\mathrm{Im}(\mathbf{A}(\hat{\mathcal{P}})\mathbf{x}+\mathbf{v})\preceq\mathrm{Im}(\hat{\mathbf{y}}^{\mathrm{up}})\middle|\mathbf{x}, \hat{\mathcal{P}}\right].
\end{align}
Again, $\ell(\mathbf{x}, \hat{\mathcal{P}})$ has a well-known closed form expression that is log-concave with respect to $\mathbf{x}$, which can be verified using the same argument from \eqref{likelihood_function}. So, NFCFGS estimates $\mathbf{x}$ according to the maximum likelihood criterion as
\begin{equation}
\hat{\bm{\alpha}}=\argmax_{\mathbf{x}\in\mathbb{C}^{|\hat{\mathcal{P}}|}}\underbrace{\log\ell(\mathbf{x}, \hat{\mathcal{P}})}_{=g(\mathbf{x}, \hat{\mathcal{P}})},
\end{equation}
whose global optimum is attained by convex optimization, gradient descent method, for example.

The NFCFGS algorithm is outlined in Algorithm \ref{nfcfgs_cv}, and the CV technique along with $\mathcal{E}$ and $\mathcal{CV}$ in the superscripts and subscripts are explained shortly after. In Line 5, $(\theta, \tau, k)$ is coarsely estimated on the grid. Then, $(\hat{\theta}, \hat{\tau})$ is refined off the grid using Newton's method in Lines 8 and 10. After updating $\hat{\mathcal{P}}$ in Line 13, the path gains are estimated using convex optimization in Line 14. The exact forms of $f_{\mathbf{z}}(\cdot, \cdot, \cdot)$, $\mathbf{n}_{\mathbf{z}}(\cdot, \cdot, \cdot)$, $\mathbf{g}_{\mathbf{z}}(\cdot, \cdot, \cdot)$, and $g(\cdot, \cdot)$ are omitted due to the page limit, and the interested reader is referred to \cite{9351751}.

\begin{algorithm}[t]
\caption{NFCFGS-CV algorithm}\label{nfcfgs_cv}
\begin{algorithmic}[1]
\Require $\hat{\mathbf{y}}$
\Ensure $(\hat{\bm{\alpha}}, \hat{\mathcal{P}})$
\State // $g_{\mathcal{CV}}(\hat{\bm{\alpha}}, \emptyset)=-\infty$ and $\mathbf{A}(\emptyset)\hat{\bm{\alpha}}=\mathbf{0}_{RN}$ by convention
\State $\hat{\mathcal{P}}\coloneqq\emptyset$
\Do
\State $\epsilon\coloneqq g_{\mathcal{CV}}(\hat{\bm{\alpha}}, \hat{\mathcal{P}})$ and $\mathbf{z}\coloneqq\mathbf{A}(\hat{\mathcal{P}})\hat{\bm{\alpha}}$
\State $(\hat{\theta}, \hat{\tau}, \hat{k})\coloneqq\displaystyle\argmax_{(\theta, \tau, k)\in\Omega}f_{\mathbf{z}}^{\mathcal{E}}(\theta, \tau, k)$
\While {termination condition}
\If {$\nabla_{(\theta, \tau)}^{2}f_{\mathbf{z}}^{\mathcal{E}}(\hat{\theta}, \hat{\tau}, \hat{k})\prec\mathbf{0}_{2\times 2}$}
\State $(\hat{\theta}, \hat{\tau})\coloneqq(\hat{\theta}, \hat{\tau})+\eta\mathbf{n}_{\mathbf{z}}^{\mathcal{E}}(\hat{\theta}, \hat{\tau}, \hat{k})$
\Else
\State $(\hat{\theta}, \hat{\tau})\coloneqq(\hat{\theta}, \hat{\tau})+\eta\mathbf{g}_{\mathbf{z}}^{\mathcal{E}}(\hat{\theta}, \hat{\tau}, \hat{k})$
\EndIf
\EndWhile
\State $\hat{\mathcal{P}}\coloneqq\hat{\mathcal{P}}\cup\{(\hat{\theta}, \hat{\tau}, \hat{k})\}$
\State $\hat{\bm{\alpha}}\coloneqq\displaystyle\argmax_{\mathbf{x}\in\mathbb{C}^{|\hat{\mathcal{P}}|}}g_{\mathcal{E}}(\mathbf{x}, \hat{\mathcal{P}})$
\doWhile $g_{\mathcal{CV}}(\hat{\bm{\alpha}}, \hat{\mathcal{P}})>\epsilon$
\end{algorithmic}
\end{algorithm}

\subsection{Proposed CV Technique}
The optimal termination condition of NFCFGS depends on $\{L_{k}\}_{\forall k}$. In practice, however, the prior knowledge on $\{L_{k}\}_{\forall k}$ is hardly available, which calls for a more feasible termination condition. In this paper, the CV technique \cite{4301267, 5319752} is adopted as an indicator of termination, which is a model validation technique that assesses the quality of the estimate to prevent overfitting. The resulting algorithm is dubbed NFCFGS-CV.

To apply CV, $\hat{\mathbf{y}}$ is partitioned to the estimation data $\hat{\mathbf{y}}_{\mathcal{E}}\in\mathbb{C}^{|\mathcal{E}|}$ and CV data $\hat{\mathbf{y}}_{\mathcal{CV}}\in\mathbb{C}^{|\mathcal{CV}|}$ where $\mathcal{E}$ and $\mathcal{CV}$ are the index sets that form a partition of $\llbracket RN\rrbracket$. Then, channel estimation is performed based on $\hat{\mathbf{y}}_{\mathcal{E}}$, while the estimation quality is assessed based on $\hat{\mathbf{y}}_{\mathcal{CV}}$. The disjoint nature of $\hat{\mathbf{y}}_{\mathcal{E}}$ and $\hat{\mathbf{y}}_{\mathcal{CV}}$ enables CV to assess the estimation quality without any bias.

To solidify the concept of NFCFGS-CV, define the estimation functions
\begin{equation}\label{estimation_function}
\begin{alignedat}{2}
&f_{\mathbf{z}}^{\mathcal{E}}(\theta, \tau, k)         &&\text{: square of the magnitude of the derivative},\\
&\mathbf{n}_{\mathbf{z}}^{\mathcal{E}}(\theta, \tau, k)&&\text{: Newton step},\\
&\mathbf{g}_{\mathbf{z}}^{\mathcal{E}}(\theta, \tau, k)&&\text{: gradient step},\\
&g_{\mathcal{E}}(\mathbf{x}, \hat{\mathcal{P}})        &&\text{: log-likelihood function},
\end{alignedat}
\end{equation}
and CV function
\begin{equation}\label{cv_function}
g_{\mathcal{CV}}(\mathbf{x}, \hat{\mathcal{P}})\text{: log-likelihood function}
\end{equation}
where the estimation (and CV) functions are obtained by deriving $f_{\mathbf{z}}(\theta, \tau, k)$, $\mathbf{n}_{\mathbf{z}}(\theta, \tau, k)$, $\mathbf{g}_{\mathbf{z}}(\theta, \tau, k)$, and $g(\mathbf{x}, \hat{\mathcal{P}})$ as in the previous subsection but under the assumption that the quantized received signal is $\hat{\mathbf{y}}_{\mathcal{E}}$ (and $\hat{\mathbf{y}}_{\mathcal{CV}}$) instead of $\hat{\mathbf{y}}$. Then, NFCFGS-CV proceeds as follows in Algorithm \ref{nfcfgs_cv}. In Lines 5, 8, 10, 13, and 14, NFCFGS performs channel estimation as discussed in the previous subsection but based on \eqref{estimation_function}. Then, Line 15 cross validates the estimation quality based on \eqref{cv_function}, which is terminated when the estimation quality starts to fall---an indication of overfitting. As a sidenote, $g_{\mathcal{CV}}(\hat{\bm{\alpha}}, \hat{\mathcal{P}})$ is in fact proportional to $-\|\hat{\mathbf{h}}-\mathbf{h}\|^{2}$, which is rigorously stated and proved in \cite{9351751}. Therefore, CV enables NFCFGS-CV to achieve the minimum squared error (SE) by terminating at the right timing.

\section{Simulation Results}
In this section, the accuracy of the NFCFGS-CV-based channel estimator is investigated based on the normalized mean SE (NMSE) $\mathbb{E}\{\|\hat{\mathbf{h}}-\mathbf{h}\|^{2}/\|\mathbf{h}\|^{2}\}$. The base station employs a half-wavelength antenna spacing uniform linear array (ULA) with uniform ADCs, whose quantization intervals vary with the SNR. The columns of the RF combiner are composed of circularly shifted ZC sequences. Likewise, the training signals are constructed as circularly shifted ZC sequences, while the raised-cosine (RC) pulse shaping filter is adopted with a roll-off factor of $0.35$. For NFCFGS-CV, $80$\% of the training signals are allocated to the estimation data, and the remaining $20$\% to the CV data.

As baselines, the GAMP \cite{8171203}, VAMP \cite{8171203}, GEC-SR \cite{8320852}, and VB-SBL \cite{8310593} algorithms are considered, which are state-of-the-art on-grid CS algorithms for low-resolution ADCs. GAMP, VAMP, and GEC-SR model the prior distribution of $\mathbf{h}$ as a Gaussian mixture, while VB-SBL uses the Student-t distribution. Unlike NFCFGS-CV, these algorithms use all the training signals to estimate the channel.

\begin{figure}[t]
\centering
\includegraphics[width=1\columnwidth]{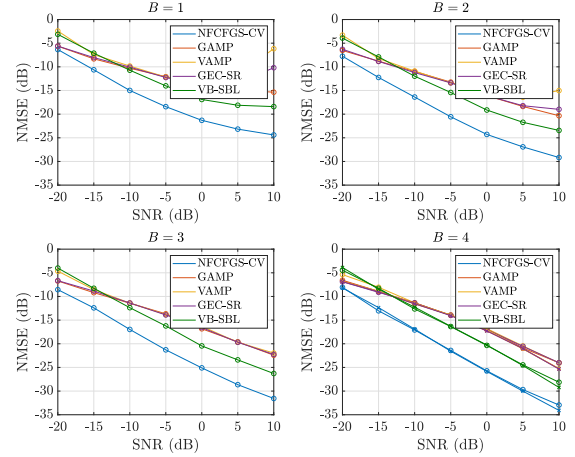}
\caption[caption]{NMSE versus SNR for $B=1, 2, 3, 4$. In the last subplot, $B=\infty$ is shown as a reference with a cross.}\label{figure_1}
\end{figure}

In Fig. \ref{figure_1}, the NMSEs of NFCFGS-CV and other baselines are shown for various SNRs with $B=1, 2, 3, 4$. The simulation setting is $M=32$, $R=8$, $K=4$, $D=4$, $L_{k}=2$ for all $k$, $N=1600$, and $T_{s}=1/600$ $\mu$s, which translates to $600$ MHz bandwidth. Fig. \ref{figure_1} shows the superior performance of NFCFGS-CV over other baselines at all SNRs and $B$. The performance gain is due to the gridless nature of NFCFGS-CV along with the proper termination condition that captures the minimum SE-achieving timing. Therefore, NFCFGS-CV yields an accurate channel estimate, which is critical to establishing a reliable data transmission link.

\section{Conclusion}
The NFCFGS-CV-based channel estimator was proposed for mmWave hybrid massive MIMO systems with low-resolution ADCs. The channel was estimated by the NFCFGS algorithm, which is a gridless CS algorithm that performs single path estimation at each iteration. The CV technique was adopted to identify the proper termination condition when the number of paths is unknown a priori. The simulation results demonstrated that NFCFGS-CV outperforms state-of-the-art on-grid CS algorithms for low-resolution ADCs.

\bibliographystyle{IEEEtran}
\bibliography{refs_all}

\end{document}